\documentclass[aps,eqsecnum,showpacs,twocolumn]{revtex4}
\usepackage{graphicx}%
\usepackage{amsmath}%
\usepackage{amsfonts}%
\usepackage{amssymb}
\usepackage{bm}
\begin{document}
\draft

\preprint{not yet sub}

\title{A chaotic 'turnstile' for atoms in periodic potentials}

\author{ M. R. Isherwood and T. S. Monteiro}

\affiliation{Department of Physics and Astronomy, University College London,
Gower Street, London WC1E 6BT, U.K.}

\date{\today}

\begin{abstract}

Using a new type of chaotic ratchet generated by pulsed standing waves of
light,  we propose a mechanism which would allow packets of
atoms travelling through a pulsed optical lattice in one direction
 to pass almost undisturbed, while strongly heating  atoms drifting
 through in the
opposite direction. An analytical formula for the diffusive energy growth
is derived and shown to give good agreement with numerical calculations.
\end{abstract}


\maketitle

There is much ongoing interest in ratchets- in other words the use of
periodic (but spatially and/or temporally asymmetric) systems which can be
used to generate a current when there is no net force: see
\cite{Reimann2} for a comprehensive review. Much of the work 
has focussed on Brownian ratchets \cite{Brown} which play an important role 
in biophysical systems such as molecular motors. Corresponding quantum ratchets
have also been investigated \cite{Reimann}.

     There has been comparatively little work on Hamiltonian ratchets,
which are 'clean' ratchets  without extrinsic noise or dissipation.
Mixed phase-space mechanisms (involving tori/stable islands)
for ratchet transport have been investigated
\cite{Flach,Ditt,Cheon}. In \cite{Flach,Flach1} general properties of ratchets 
and a mechanism involving desymmetrization of ballistic flights was
proposed. In \cite{Ditt} a sum rule connecting the current in  
regular islands with that in adjoining chaotic manifolds was derived.

Until recently it was not thought that a fully chaotic system
could  admit a ratchet effect. But in \cite{Mon1} 
 a fully chaotic ratchet mechanism was 
proposed for atoms in a double-well periodic lattice, pulsed with unequal
periods. The ratchet depends  on the generic properties
of the chaotic diffusion so the initial state requires no 
specific preparation: in all cases it was found that
in general there is a classical timescale
$t_r$, during which particles moving with negative momenta (relative to the
initial value) absorb energy at a different rate from those moving in
the opposite direction. In the corresponding quantum ratchet, for a significant
effect to develop, it was found in \cite{Mon1} that it is simply necessary that 
the quantum break-time  $t^*$ be of similar magnitude to $t_r$.

Atoms in pulsed optical lattices have become a paradigm in the study of
quantum chaos; experiments on sodium and cesium atoms \cite{Raizen}
have provided a convincing demonstration of dynamical localization (DL)
\cite{Casati,Fish},
 the quantum suppression of chaotic diffusion. The dynamics for the usual
experiment is given by the kicked-particle Hamiltonian :
 $H=\frac{p^2}{2} -  K \cos x \sum_n \delta(t-nT)$
where $K$ is the kick strength. The classical dynamics is obtained by
iterating the well-known 'Standard Map' :
$x_{i+1}=x_i + p_i T$; $ p_{i+1}= p_i + K \sin x_{i+1}$. 

We can take $T=1$, without loss of generality, in the Standard Map, but
for the ratchet, we use a repeating cycle of unequal kicks.
 The ratchet Hamiltonian is given by
 $H=\frac{p^2}{2} +  V(x) \sum_{n,j} \delta(t-nT_j)$
where for the $n-$th  cycle, the period of the $j-th$
kick is $T_j$. For example, in \cite{Mon1} the kick spacings
cycle between $T_1=1+b, T_2=1, T_3=1-b$ where $|b| <1$.

 In \cite{Mon1} an asymmetric potential of the
form $ V(x)=  K \{ \sin x + a \sin (2x + \Phi) \}$
was investigated, but in \cite{Cheon} the spatial
symmetry was broken by a 'rocking' linear term:
$V(x)=-(K \cos x + Ax s_j)$ where $s_j=(-1)^j$.
In \cite{Cheon} the kick spacing alternated between
$T_1=1+b$ and $T=1-b$.

The rocking ratchet \cite{Cheon}
was  analysed in the regime where regular tori are still
present ; it was found that although it
no longer has the $2 \pi$ periodicity in $p$ of the Standard Map,
there is a long range periodicity in $p$ which is of order $2\pi/b$.
Also, since pairs of corresponding tori are located asymmetrically 
about $p=0$, the resulting confinement of classical trajectories
can yield transport.

Here we investigate this system in the chaotic regime.  
One aim of this work is to demonstrate
that the physics is similar to that
found in \cite{Mon1} and hence that the new chaotic ratchet mechanism is quite
generic: both
 ratchets depend on the ratchet timescale
  $t_r \sim \frac{1}{b^2D}$ in the classical case, where
$D$ is the diffusion rate, and the break-time $t^* \sim D/ \hbar^2$ in the
quantum case. The ratchet currents were found in \cite{Mon1}
 to originate predominantly from correlations of the form
 $<V'(x_i)V'(x_{i+2})>$ . We show below that the resulting 
analytical formulae accurately
predict phenomena such as current reversals, without any detailed consideration
of the structure of phase-space.

Further, while \cite{Mon1} considered only the ratchet current which arises
for a system with  zero initial current, ie $<p(t=0)>=0$, here we
present results for the case where there is an initial drift $<p(t=0)>=p_0$.
We show that this can form the basis of a device to manipulate traffic
of cold atoms moving along some channel in a trap. This is not so
far-fetched, since the trapping of atoms in devices such as atom
 chips is now a reality. The proposed device represents a sort of
 'turnstile'  which would 
selectively heat only atoms moving in one direction, while imparting
little or no energy to atoms moving in the opposite direction, since we
can tune the parameters of the pulsed lattice so as to generate
 correlations which almost cancel the energy diffusion
in one direction, but enhance it in the other. Since the rocking ratchet
produces much simpler analytical expressions than the double well ratchet, 
it provides a better didactic example of the underlying physics.

For the Standard Map , at the lowest level of approximation,
the momenta at consecutive kicks are uncorrelated and evolve in time as a
'random-walk'. The average momentum of a large ensemble of particles is unchanged.
The average energy grows linearly: if the momenta are uncorrelated the average
kinetic energy grows by $K^2/4$ at each consecutive kick. 
In the absence of phase-space
barriers the energy is unbounded and this diffusive increase in energy 
 continues indefinitely. It is characterized by a diffusion rate $D_0$, ie 
$<p^2> =D_0 t$ so for uncorrelated momenta
$D_0=K^2/2$. However, this results neglects correlations between 
 sequences of consecutive kicks; if included, they result in
 well-known corrections to the
diffusion constant in the form of Bessel functions hence
 $D_0 = \frac{K^2}{2}(1- 2(J_1(K))^2-2J_2(K)...)$
\cite{Lich,Shep}. These corrections have even been measured experimentally
with cold cesium atoms in pulsed optical lattices \cite {Raizacc}.

Here, the $C(2)= -K^2 J_2(K)$ term is of particular interest.
 It corresponds to a
two-kick correlation, given by evaluating the phase-space average of
$2<V'(x_{i-1}) V'(x_{i+1})>$ hence:
\begin{equation}
 C(2) =2K^2 < \sin x_{i-1} \sin x_{i+1}>.
\end{equation}
It is easy to show, by direct substitution
from the map that one obtains separate integrals over the $x$ and $p$ .
For chaotic diffusion it is usual to neglect odd integrals,
for example $ <\sin p \cos p> \simeq 0$, while 
even integrals yield a positive average, eg $<\sin ^2 p> =<\cos^2 p>=1/2$,
since after a few kicks the ensemble has a substantial spread in $p$.
In effect the only terms which contribute to $C(2)$ in the standard
map are even in $p$ and do not differentiate between $ \pm p$.
The integral over $x$ yields $J_2(K)$, hence the form of $C(2)$.

We consider now the map of \cite{Cheon},
 with a rocking linear potential, which consists of a
repeating cycle of two kicks. Around the $i-th$ kick we have:
\begin{eqnarray*}
x_i = x_{i-1} + p_{i-1} (1+b)\\ 
p_i = p_{i-1} + K \sin x_i + A \\
x_{i+1} = x_i + p_i(1-b)\\
p_{i+1} = p_i + K \sin x_{i+1} - A\\
\end{eqnarray*}
Following the usual procedure, we now obtain a modified form for
the 2-kick average, but find that the momentum averages now include
terms of the form $<\sin^2p. \sin 2pb>$ and $<\cos^2p. \sin 2pb>$.
 At small $t$, $pb$ is small while $\sin^2 p$ oscillates rapidly, hence we
can approximate these by 
$1/2<\sin 2pb> \sim p_{av}(t) b$ and $1/2<\cos 2pb> \sim 1/2$ where $p_{av}$ is
the average momentum relative to the intial momentum.
 We note that such terms are the
origin of the ratchet effect proposed in \cite{Mon1}. 
These terms are odd in $p$
and hence- unlike the standard map case-
 depend on whether we average our momenta from $0 \to \infty$ or from
$0 \to -\infty$. As in \cite{Mon1} below we consider  the 
case of particles with positive momenta separately from
 those with negative momenta,
since the chaotic ratchet depends on differential energy diffusion rates
for particles moving in opposite directions.

We consider first the case where 
$p(t=0)=0$ (or we take a cloud of particles with a 
narrow gaussian distribution peaked about $p=0$),
 but later consider the case where the initial current is
non-zero. We  take $p_{av} \sim \pm \sqrt{D_0 t}$ for particles
with positive or negative momenta respectively; hence $|p_{av}|$ estimates the
width of the momentum distributions about the origin.  

With the approximations $<\cos 2pb> \simeq 1$ and $<\sin 2pb> \simeq 2pb$
  it is easy to show:
\begin{equation}
C^{\pm}(2) \simeq -K^2 J_2(K)[ \cos A(1 \pm b) +2p_{av}(t)b \sin A(1 \pm b)]
\end{equation}

The $\pm$ correspond to the correlations between the two possible 
sequences of two kicks
in the map, and we must average these to obtain $C(2)=1/2[C(2)_{-} + C(2)_{+}]$.
However, since $b$ is usually a very small parameter, here we can take:
\begin{equation}
C(2) \simeq -K^2 J_2(K)[\cos A + 2p_{av}(t) b \sin A]
\end{equation}
or considering separately the energy diffusion for positive
 and negative momenta:
\begin{equation}
C(2,\pm) \simeq -K^2 J_2(K)[\cos A  \pm 2 \sqrt{D_0t}  b \sin A]
\label{c2}
\end{equation}

We see immediately from this expression that only the terms in $\sin A$
 differentiate between positive and negative momenta and hence allow a
ratchet effect. This implies that for $A = n \pi$, where $n=0,1,2..$
is an integer, there is no transport, while for $A = (2n+1) \pi/2$, there
is maximal transport.
 
\begin{figure}[ht]
%\vskip -1.5in
\includegraphics[height=2.5in,width=3.in]{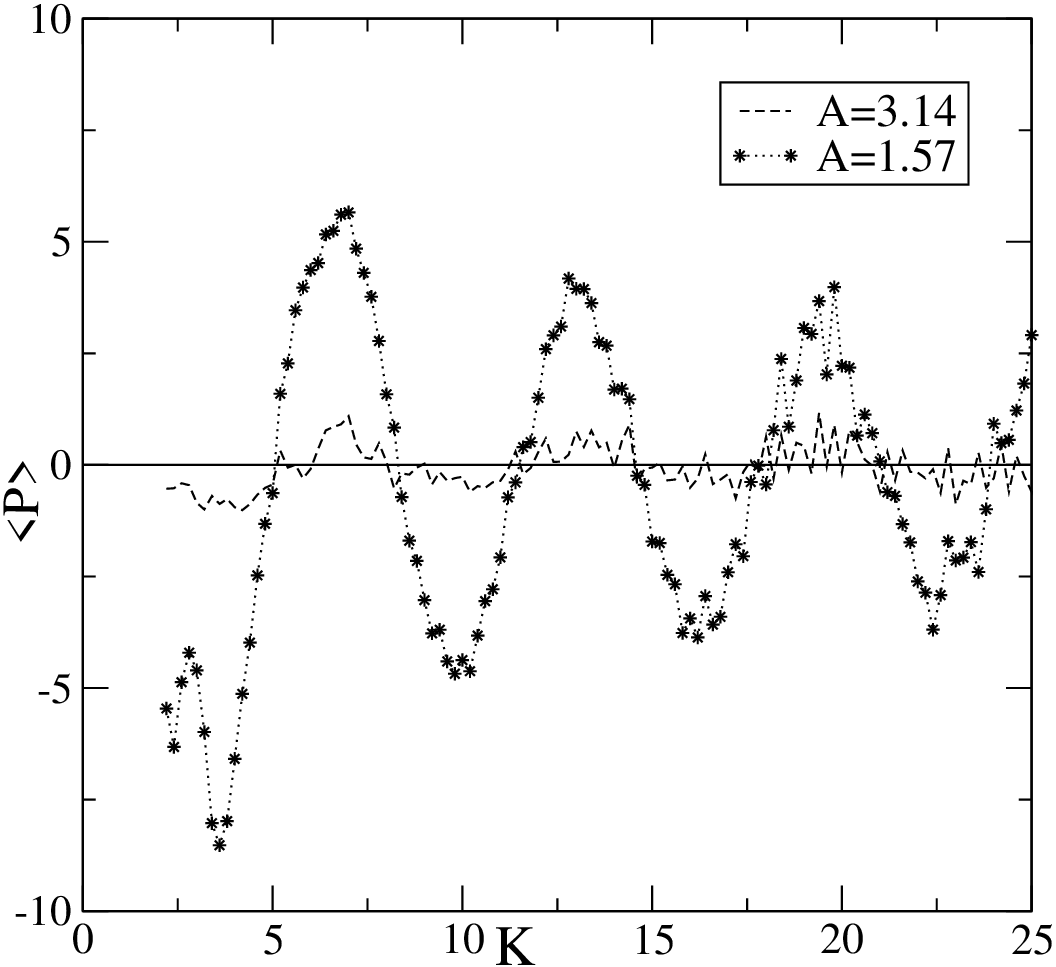}
%\vskip 0.5in
%\includegraphics[height=1.5in,width=3.in]{fig1b.eps}
\caption {Figure compares the behavior of the classical current as a function
of kick strength $K$ for $A \simeq \pi/2$ , $b=0.03$ and 
$A \simeq \pi$. It shows that the latter case is not transporting (ie $<P> 
\simeq 0$), but that for $A= \pi/2$ the current direction follows the
oscillations of a Bessel function $-J_2(K)$.
 }  
\label{Fig.1}
\end{figure}

In Fig.1 we calculate numerically the classical current, as a function
of $K$, for an ensemble
of 300,000 trajectories initially sharply peaked about $p=0$ (ie a gaussian
distribution of width $\sigma=1$). For $A \sim \pi$ there is essentially no
current, while for $A \sim \pi/2$, the current is appreciable and
current reversals follow exactly the oscillations of $-J_2(K)$, as 
expected from the formula.  

In \cite{Mon1}, the origin of the classical current 
was also found to be due to the fact that
for $p_{av} >0$ we have a $C(2)$ correction of the opposite sign to the
correction for $p_{av} <0$, though in the case of the double-well
ratchet the analytical form
of the correction is much more complicated than Eq.\ref{c2}.

In Fig.2 we test Eq.\ref{Energ} against a numerical calculation of the
energies as a function of  time for $K=3$ and $K=10$ and $A=\pi/2$.
 For an ensemble of trajectories initially peaked
about $p=0$, we expect that
 the average energy,$E_{+}$, of particles with positive momenta,
 and the average energy,
$E_{-}$, for those with negative momenta, are given by:
\begin{equation}
2E_{\pm} \sim D_0 t \pm \frac{4}{3}b K^2 J_2(K) \sqrt{D_0}  t^{3/2}
\label{Energ}
\end{equation}  
in the regime where $pb$ is small. For large $pb$, 
$<\sin 2pb> \simeq <\cos 2pb> \simeq 0$ hence 
$C(2)=0$; as shown in \cite{Mon1}, beyond a time scale
$t_r$ that it takes the momenta to increase to $|p_{av} b| \sim 1$
both the positive $p$ and negative $p$ part of the ensemble of trajectories
diffuse at the same rate $2E=<p^2>=D_0 t$, where for this system 
$D_0 \simeq K^2/2 (1-2J_1(K)^2..)$. Hence $t_r \sim 1/(Db^2)$.

 The graph shows that for short
times, the equation gives a good estimate to the correction to the
 energy,($<p^2>-D_0t$).
It also shows that after $t \sim t_r$ the correction vanishes, and
 $<p^2>-D_0t$
becomes a constant. 

\begin{figure}[]
%\vskip -1.5in
\includegraphics[height=2.in,width=3.in]{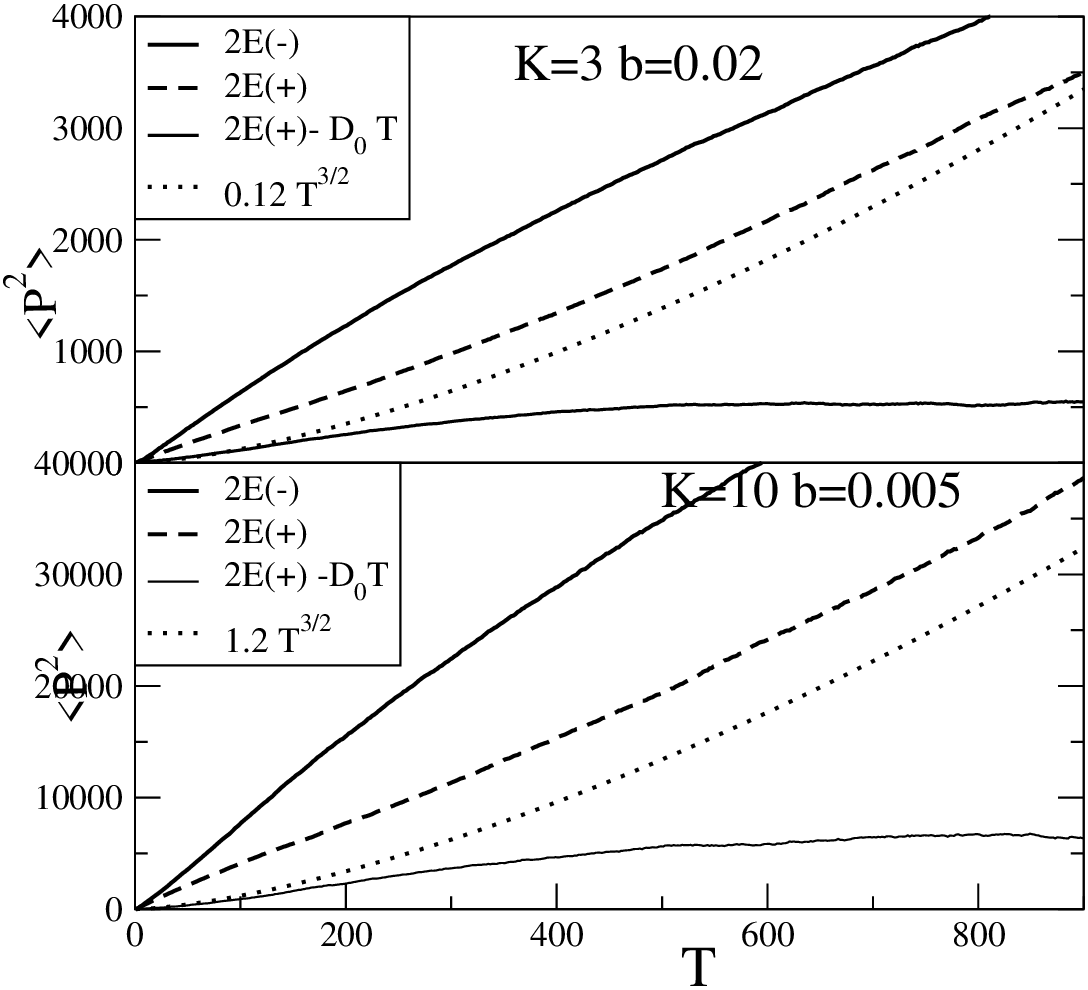}
\caption {Comparison between average energy of particles with positive
momenta $E(+)$ and those moving with negative momenta $E(-)$ for
two different $K$ and $b$. The figure shows that the correction to
the linear energy growth is well described at short times by 
Eq. \ref{Energ}; at longer times the ratchet correction vanishes and
$E(\pm)$ grow as $\sim D_0t$. }  
\label{Fig.2}
\end{figure}

It is very interesting to consider also the case where at $t=0$ we already
have a drift current; in other words we start our trajectories with
non-zero $p=p_0$. It is easy to show that the corresponding correction to the
energy diffusion,$C(2,p_0,p^{\pm}_{av})$, now takes
the form:
\begin{eqnarray*}
 \simeq -K^2 J_2(K)
 \{ \cos A(\cos 2p_0b + 2p^{\pm}_{av}(t)b \sin 2p_0b)\\ 
                + \sin A(\sin 2p_0b + 2p^{\pm}_{av}(t)b \cos 2p_0b) \}
\label{diff}
\end{eqnarray*}
$p^{\pm}_{av}(t)$ now represents the average momenta relative to $p_0$
at time $t$
for particles with momenta greater or less than $p_0$ respecively.
 We see that the average energy absorption rate of the cloud of atoms  
depends on $p_0$ as well the the effective widths of
the distribution $p^{\pm}_{av} \sim \pm \sqrt{D_0t}$ about $p_0$.
In other words, initially, if all particles are at $p=p_0$ , we have a local
diffusion rate correction $-K^2 J_2(K) \cos (A-2p_0b)$; but as the
cloud spreads in phase space one has to consider the average over the width 
of the cloud.

When $pb$ is large, the $\cos 2pb$ and $\sin 2pb$ terms
average to zero, hence $C(2,p_0) \to 0$ asymptotically for $t>t_r$ regardless
of the value of $p_0$. After this, the average
energy growth for the particles with positive
 and negative momenta (relative to $p_0$)
is equal to $D_0t$. Regardless of $p_0$, the timescale for
these correlations to be 'averaged-out'
is given by the ratchet time of \cite{Mon1},
$t_r \sim (D_0b^2)^{-1}$, the time required for the 
distribution to broaden to a substantial width in $pb$.
In \cite{Cheon} it was found that although the classical map  does not have
the $2 \pi$ periodicity of the standard map, it has a new 
long-ranged periodicity in $p$, with 
periodicity  $\sim 2 \pi/b$. From the formula, we see
that the periodicity of the correction is half of this:
 $C(2, p_0=0) = C(2, p_0= \pi/b)$.
We note that since the $C(2)$ correction vanishes when the distribution
size becomes of order one cell in $p$, the long range periodicity 
in this respect is not significant to
the magnitude of current accumulated:
by the time the expanding distribution samples the periodicity boundary,
$<p>$ is almost constant: we have tested this numerically.

We now consider how one might exploit Eq.\ref{diff} to construct a
cold atom 'turnstile'. For simplicity we take the case $A= \pi/2$.
Consider the case of a packet of cold atoms drifting through the lattice
with a constant drift, which may be positive or negative,
 $p_0b=\pm \frac{\pi}{4}$; we see that for short times
(where the equation is valid) its energy grows
linearly, so $<p^2> \simeq Dt \simeq [D_0 - (\pm K^2 J_2(K))]t \simeq 
\frac{K^2}{2}[1- 2(J_1(K))^2 - (\pm 2J_2(K))]t$. In other words
for positive drift, the energy absorption is reduced relative to
the basic diffusion term, while for negative drift it is accelerated.
For $ K \sim 2$ we can cancel the diffusion term almost completely
so for $p_0b =\pi/4$ the packet absorbs very little energy.
\begin{figure}[ht]
%\vskip -1.5in
\includegraphics[height=2.5in,width=3.in]{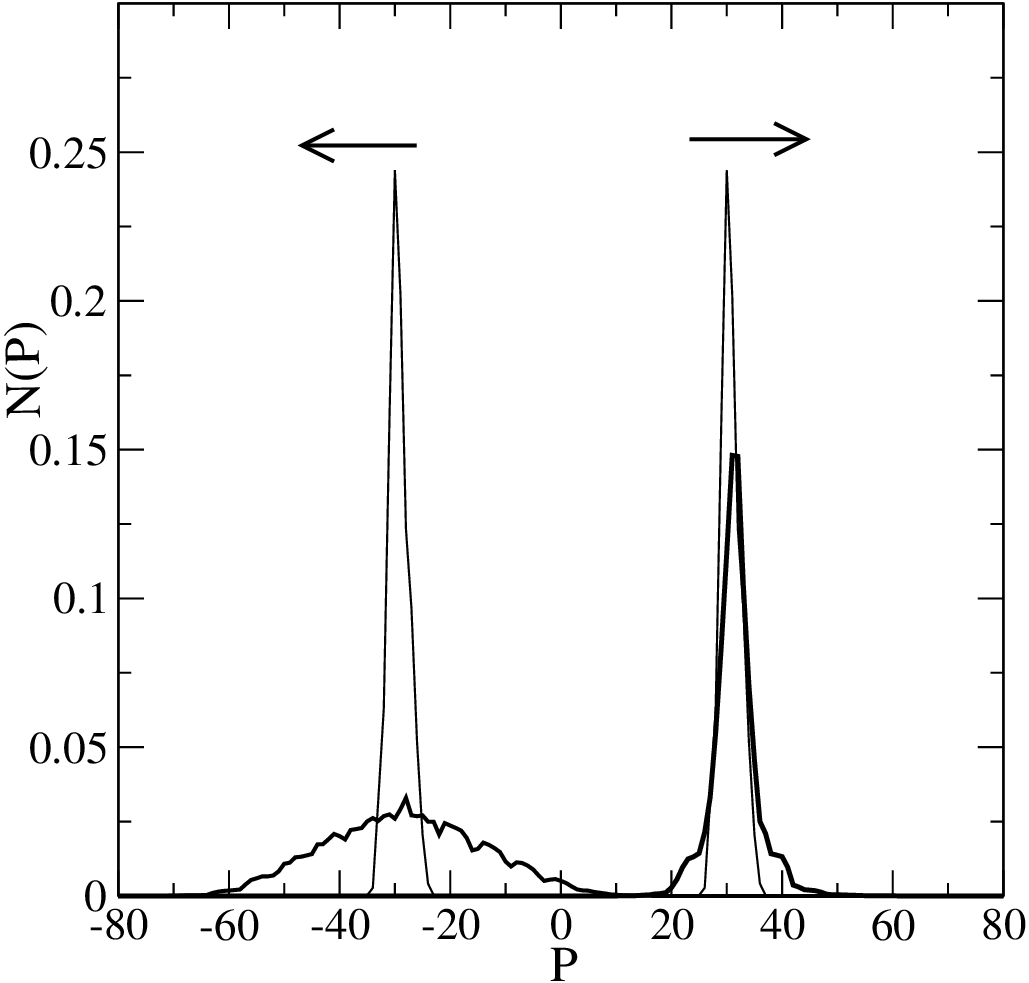}
%\vskip 0.5in
\includegraphics[height=2.5in,width=3.in]{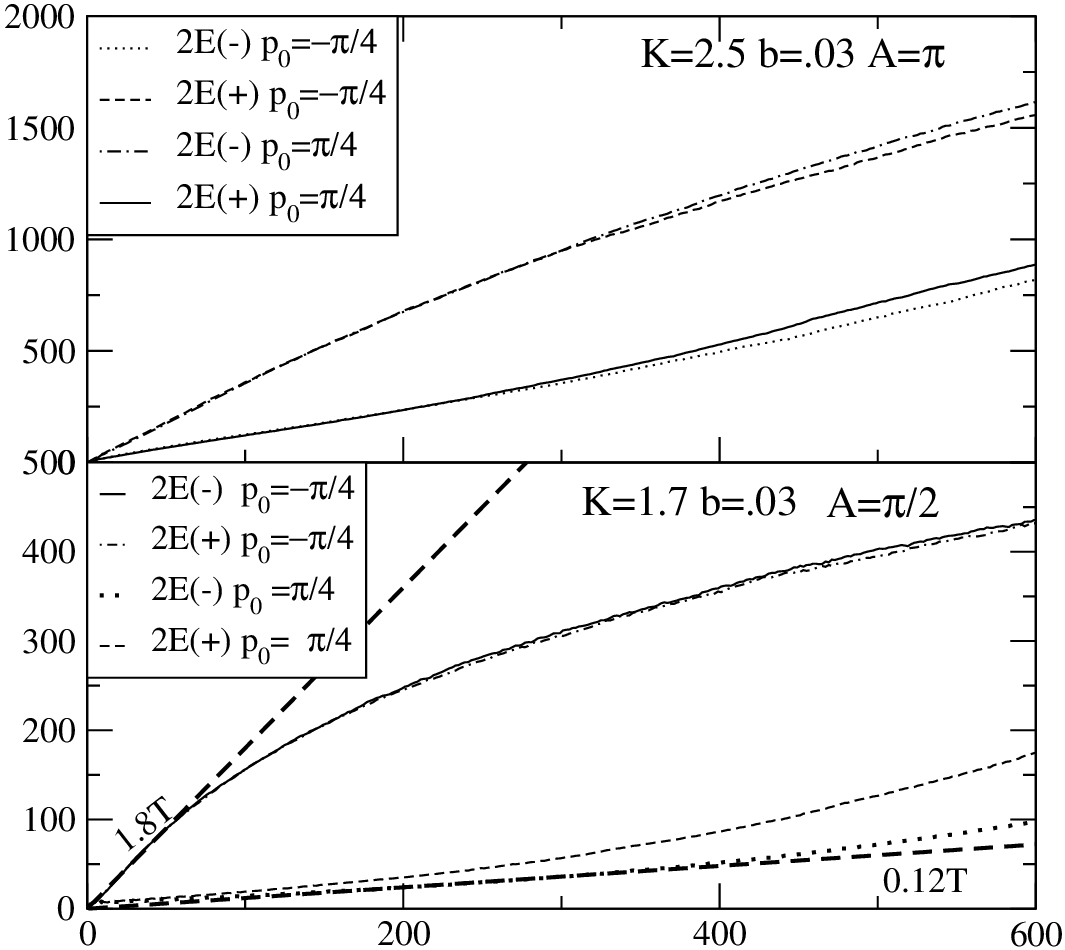}
\caption {Shows the effect of a pulsed lattice on a system
with finite initial current.
Figure 3a demonstrates the 'turnstile' effect on particles 
moving through the lattice with initial momenta
 $p_0 = \pm 27 \simeq \pi/(4b)$.
The initial momentum distributions are shown, as well as the effect of
$\simeq 125$ kicks on the final distributions. We see that while
 particles moving right have absorbed relatively little energy, 
the particles moving left have substantially increased average kinetic
energy.
In Fig 3b we show that Eq.\ref{diff} accurately describes the energy absorption 
at short times. The upper figure shows energy growth for $\cos A= \pi$ and
$p_0b= \pm \pi/4$. Energy absorption is asymmetric about $p_0$ and is greater
than $D_0t$ if $p_0, p_{av}$ are of different sign, but smaller otherwise.
The lower figure corresponds closely to Fig3a. Energy absorption rates
are symmetric about $p_0$ but are $\simeq 15$ times faster for negative
momenta relative to positive momenta. }  
\label{Fig.3}
\end{figure} 

Fig.3a shows the effect  on two classical ensembles drifting through the ratchet 
with speeds of $p_0=\pm 27$
(for an effective turnstile it is not  essential
to have exactly $p_0=\pm \pi/(4b) $) and $b=0.03, K=1.7$. Here $D_0 \simeq 0.8$
so for positive $p$, $D \sim 0$, while for negative $p$
$D \sim 1.6$. The figure shows the initial distributions as
well as the average momentum distribution after $\sim 125$ kicks. We see that the
effect of the pulsed lattice on the two components is drastically different.
While the cloud moving right is only slightly perturbed, the  cloud
moving left
has been heated to much larger average kinetic energies.
 Since for this case $C(2,p_0,p^{\pm}_{av})$ now  has no dependence on $p_{av}$
for short times,
the particles absorb energy symmetrically about $p_0= \pm \pi/(4b)$.

The corresponding quantum case was investigated in 
\cite{Mon1}; for the equivalent quantum turnstile to show similar behaviour
we require only that the quantum break-time $t^*$ should be of the same order 
or longer than the duration of the classical turnstile (ie $\sim 100$ kicks
for the above example, though we can adjust this by varying $b$).
For $t < t^*$, the quantum behaviour follows closely the classical
behaviour. For $t>t^*$, the quantum wavepacket localizes and absorbs
no more energy, thus 'freezing-in' the ratchet asymmetry.
For the turnstile, if the lattice region is of finite extent, the
limiting time could be the time the atom spends within the
optical lattice if this is less than $t^*$.

In Fig.3b we further verify Eq.\ref{diff} by plotting the actual energy growth 
for the case $A=\pi$ (upper graph) and $A= \pi/2$ (lower graph). 
For the upper graph, only the $\cos A$ terms contribute; since we 
take $p_0= \pm \pi/4$, we can see from both the formula 
($C(2,p_0,p^{\pm}_{av}) \simeq - 2bp^{\pm}_{av} \sin 2p_0b $) and the numerics
that for $p_0,p_{av}$ of the same sign, trajectories absorb energy 
slower that $D_0t$ while for $p_0,p_{av}$ of different sign the converse
is true. The lower figure corresponds closely to the 'turnstile' 
shown in Fig.3a since we have $\sin A= \pi/2$. The figure shows that
for short times, the energy of particles initially with 
$p_0= \pm \pi/4$ grows linearly. However the negative momentum particles      
absorb energy $ \simeq 15$ times faster than the particles with a
positive drift.

In conclusion, we have shown the chaotic ratchet effect found in \cite{Mon1}
is generic in character and applies to another class of ratchets.
We show that by means of a simple analytical formula the effect can 
be maximised and use this to propose a mechanism for manipulating cold atoms.
In particular we show that we can have a pulsed optical lattice that after
$\sim 100$ kicks has imparted considerable energy to particles moving in one
direction, while particles moving in the opposite direction absorb little energy.
Since the mechanism is based  on fully chaotic dynamics rather than a regime
with stable islands, no position is special so no preparation of the initial
state is required.

T.M. thanks Thomas Dittrich, Sergej Flach and
Holger Schanz for helpful discussions. M.I acknowledges an
EPSRC studentship.The work was supported by EPSRC grant GR/N19519.

\end{document}